\documentclass[11pt]{article}
\usepackage[utf8]{inputenc}
\usepackage{tabularx,lscape,longtable}
\usepackage[english]{babel}
\usepackage[dvipsnames]{xcolor}
\usepackage{amsfonts,comment}
\usepackage{multicol}
\usepackage{amssymb,latexsym,amsmath}
\makeatletter \oddsidemargin 0in \evensidemargin 0in \textwidth
16cm \RequirePackage[dvips]{graphicx} \textheight 20cm
\DeclareGraphicsExtensions{.jpg, .eps}
\newtheorem{p1}{Proposition}[section]

\definecolor{cadmiumred}{rgb}{0.89,0.0,0.13}

\begin{document}
\title{\bf
Use of Lambert $W$ function in the simulation of Weibull and non-Weibull distributions}
\author{Subhashree Patra$^{1}$, Subarna Bhattacharjee$^{2}$$^{}$\thanks{Corresponding author~:
E-mail ID: subarna.bhatt@gmail.com} \\
{\it $^{1,2}$ Department of Mathematics,Ravenshaw University, Cuttack-753003, Odisha, India}\\
}
\date{\today}
\maketitle
\begin{abstract}
In this work, we have taken up some distributions, mostly  Weibull family, whose quantile functions could not be obtained using the traditional inversion method. We have solved  the same quantile functions by using the inversion method only, with the additional help of transcendental functions like the Lambert W function. The usage of the Lambert W function has helped derive closed-form solutions for mathematical models in many fields, for which explicit or exact solutions were not available and approximation was the only known way of approaching it.\\
\end{abstract}

\section{Introduction}

${}$\hspace{1cm}Quantile functions have been extensively used for simulation and various mathematical models required in physical and financial systems. One can simulate any non-uniform random variable by applying its quantile function to uniform deviates. The use of quantile functions in Monte Carlo simulations has been discussed in Steinbrecher and Shaw (2007)

${}$\hspace{1cm}The non-availability of quantile functions in closed forms restricts its applications. The inversion method has not been successful in providing closed-form solutions due to the intractability nature of some mathematically complex cumulative distribution functions. To state just a few of them, we have distributions like normal, beta, Erlang, MacDonald, Chi, Levy, hyperbolic, beta prime, Behrens Fisher, Nakagami, Pearson, chi-square, gamma, student’s $t$, and $F$ distribution. See Okagbue et al. (2019)
\\
${}$\hspace{1cm}The most widely used alternative to obtain the quantile function of such distributions is using approximation techniques such as functional approximation, numerical algorithms, and the use of differential equations and series expansions. Numerical techniques like root finding and interpolation have been used by researchers to approximate the quantile function in  Dagpunar (1989) , Derflinger et al (2009, 2010), Lai (2009), Farnum (1991), Leydold and Horman (2011). Runge Kutta methods and Ordinary differential equations have been used as tools in Ulrich and Watson (1987), and Leobacher and Pillichshammer (2012). This aspect has been well discussed in Munir (2013).  
The simultaneous use of differential equations and series methods in analyzing quantile functions has also gained the interest of many researchers. Series expansion methods for quantile approximations broadly include Cornish Fisher expansion, orthogonal expansion, and quantile mechanism, see Steinbrecher and Shaw (2007).  

${}$\hspace{1cm}In this work, we have taken up some distributions, mostly  Weibull family, whose quantile functions could not be obtained using the traditional inversion method. We have solved  the same quantile functions by using the inversion method only, with the additional help of transcendental functions like the Lambert W function. The usage of the Lambert W function has helped derive closed-form solutions for mathematical models in many fields, for which explicit or exact solutions were not available and approximation was the only known way of approaching it.\\
${}$\hspace{0.5cm}We organize the paper as follows. Section 2 gives a brief genesis/review on Lambert $W$ function and its use in simulation process of some well-known statistical distributions whose inverse cumulative distribution ($icdf$) are not in closed form. In Section 3, we place a review on inverse probability transform of some Weibull distributions having two or more parameters giving their respective $icdf$, thereby fostering their simulation and resulting sampling procedure.  Among the Weibull distributions discussed in the text, some have $icdf$ in closed form and some Weibull family do not. Further, we go along to illustrate the significance of Lambert $W$ function in simulation of Weibull models whose $icdf$ s are not in closed form. In Section 4 we focus on some non-Weibull distributions and approximate their $icdf$ s with the help of Lambert $W$ function. 
\section{Genesis of Lambert $W$ function and its relevance in distribution theory}
${}$\hspace{1cm}We begin this section by giving a brief representation involving mathematical rigor in Lambert $W$ function followed by its relevance in statistical analyses.\\
${}$\hspace{1cm}The Lambert W function, represented by W(z) is defined as the inverse of function $ f(z) = z e ^z,$  satisfying
    $$ W (z) e ^ { w(z)} = z $$
The Tree function or Euler T function T(z) satisfies
  $ T(z) = z e ^ {T(z) }  $
 It is related to the Lambert W function by $$ T(z) = - W (-z).$$

${}$\hspace{1cm} Lambert W function is also called Omega function or the product logarithm. It is a multivalued function displaying two real branches.
The multiple branches of W are denoted by $W_k$ where $K = 0, \pm 1, \pm 2, \pm 3, \ldots,$ $W_0$ is the principal branch while $W_{-1}$ is the only other branch which takes real and unique values in some real interval. On the contrary, all the other branches are complex and multi-valued. For real $x$ lying between $\frac{-1}{e}$ and $0$, $W(x) \geq -1$ represents the branch $W_0$ whereas $W(x) \leq -1$ represents the branch $W_{-1}$ We refer \cite{corless}  and \cite{belkic} for detailed analysis of the function.  The power series representation of Lambert $W(\cdot)$ is given in the form
\begin{eqnarray}
 W(x) &=& \equiv \sum^\infty_{n=1}  \frac{(-n)^{n-1}}{n !} x^n  \nonumber\\
 &=& x - x^2 + \frac{3}{2}x^3 - \frac{8}{3}x^4
+ \frac{125}{24}x^5 - \frac{54}{5}x^6 + \frac{16807}{720}x^7 - \ldots\nonumber
\end{eqnarray}
${}$\hspace{1cm}In literature, many authors have discussed the use of the Lambert W function in various areas. We mention a few where the function has been used in finding the quantile functions.
Gayan Warahena-Liyanage and Mavis Pararai have worked on various distributions in the Lindley family. In Liyanage and Pararai(2014), they worked on the properties of exponentiated power Lindley (EPL) distribution and used the Lambert W function to find its quantile function. The special sub-models discussed are Power Lindley distribution, exponentiated Lindley (EL) distribution,  Lindley (L) distribution, an exponentiated two-component mixture of Weibull distribution and gamma distribution, and an exponentiated two-component mixture of Rayleigh distribution and a gamma distribution.
A new Lindley power series (LPS) class of distributions is introduced in Liyanage and Pararai (2015) with the Lindley binomial (LB) distribution, Lindley geometric (LG) distribution, Lindley Poisson (LP) distribution, and Lindley logarithmic (LL) distribution discussed as special cases. The quantile function is derived in closed form by directly applying the negative branch of the Lambert W function. The expression thus obtained is used to generate random data.
Similar work is done for Exponentiated Power Lindley Poisson Distribution in Pararai et al. (2016).\\
${}$\hspace{1cm}In this paper, we are particularly interested in distributions of the Weibull family. Distributions like three-parameter Weibull \cite{laixiemu}, five-parameter Weibull \cite{phani}, three-parameter Weibull \cite{xi1}, additive Weibull, Nadarajah and Kotz Weibull \cite{nada}, generalized modified Weibull and Kumaraswamy modified Weibull distribution are some of the Weibull distributions that do not have quantile functions in closed form.
Several programming languages like Fortran, C++, Matlab, Maple, Macsyma, and Mathematica (also R) have packages on Lambert $W$ function in their libraries, which facilitate the computational applications. \\
\section{Simulation of some Weibull distributions }
${}$\hspace{1cm}In this section, we follow the usual convention for some of the standard aging functions that are commonly used by reliability practitioners. By referring to aging function, we mean functions viz., reliability/survival ($SF$), hazard rate ($FR$), reversed hazard rate ($RFR$), and aging intensity ($AI$) denoted by $\bar{F}(\cdot)$, $r(\cdot)$, $\mu(\cdot)$ and $L(\cdot)$ respectively for a non-negative continuous random variable $X$. We write $X\sim N(a_1,a_2,\ldots,a_n)$ to represent that $X$ follows a certain distribution, say $N$ having parameters $a_1,a_2,\ldots,a_n.$
\subsection{Weibull distributions having Quantile function in closed form}
\subsubsection{Two parameter Weibull Distribution}
\begin{enumerate}
\item[($i$)] A  Weibull distribution (Weibull, 1951), denoted by $X\sim W_2(a,b)$ has
$SF$ $\bar{F}_X(t) = \exp { \bigg[ (-at^b) \bigg] }$ and its
inverse cumulative distribution function is given by
\begin{equation} \nonumber
   t = \bigg[ \frac{\ln(1-u)}{(-a)} \bigg] ^{\frac{1}{b}}.
\end{equation}
   \item[($ii$)] A  Weibull distribution (Gompertz, (1825)) denoted by
    $X \sim GW(a , b)   $ where $ a >0 , t \geq 0  , b \in \mathbb{R} $
   has
    $SF,$ $\bar{F}_{X}(t) = \exp { \bigg[  \frac{a}{b}  \Big( 1 - \exp (b t )    \Big)     \bigg]       }$
    and its
    inverse cumulative distribution function is
\begin{equation} \nonumber
  t = \bigg(\frac{1}{b} \bigg) \ln \bigg[1 - \frac{b}{a} \ln (1-u) \bigg].
\end{equation}

\item[($iii$)]{Truncated Log Weibull Distribution (cf. Giri et al. (2023)) }
denoted by $X \sim TLW(a,b)$ where $ b  >0 , a, t \in \mathbb{R} $
    has $SF$
 $\bar{F}_{X}(t) = \exp { \Bigg( -\exp {\Big(\frac{t-a}{b}  \Big)}\Bigg) }$
   and its
    inverse cumulative distribution function is
\begin{equation} \nonumber
    t = a + b \ln \Big( - \ln (1-u) \Big).
\end{equation}
\item[($iv$)]Flexible Weibull Distribution, denoted by (cf. Bebbington, Lai, and
Zitikis (2007)) has $SF$
$\bar{F}_{X}(t) = \exp \{ - e ^{a t - \frac{b}{t}}     \}    $
   and its Inverse Cumulative distribution Function is
\begin{equation} \nonumber
    t= \frac{- \ln  [ - \ln (1-u) \pm \sqrt{ ( \ln ( - \ln (1-u)))^2 + 4 a b   }] }{2 a}
\end{equation}
\item[($v$)]{Pham's Weibull Distribution (cf. Pham (2002))}
 $X \sim  P(  a, b )  $ where $ a >1 , b > 0,  t > 0 $ $SF$: $\bar{F}_{X}(t) = \exp (1- a^{t^b} )$
    Inverse Cumulative distribution Function:
\begin{equation} \nonumber
 t=    \bigg[  \frac{\ln [ 1- \ln (1-u) ]}{\ln a }       \bigg] ^ \frac{1}{b}
\end{equation}
\end{enumerate}
\subsubsection{Three parameter Weibull Distribution}
\begin{enumerate}
    \item[($i$)] { Exponentiated Weibull distribution (Mudholkar and
Srivastava (1993))}
    Notation used: $X \sim MS(  a, b,  c )  $ where $ a,  b, c >  0 , t > 0 $
     $SF$: $\bar{F}_{X}(t) = 1 -  [ (1- \exp (- a t^b )  ] ^ c $
       Inverse Cumulative distribution Function:
\begin{equation} \nonumber
    t= \bigg[ \bigg(-  \frac{1}{a}  \bigg)   \ln (1-u^\frac{1}{c} ) \bigg] ^ \frac{1}{b}
\end{equation}
    \item[($ii$)]{Modified Weibull Extension Distribution (cf. Xie, Tang, and Goh (2002))}
    Notation used: $X \sim XTG ( a,  b,  c)$ where $ b, a, c  >0 , t \geq 0 $
      $SF$: $\bar{F}_{X}(t) = \exp \Big\{ a b (1- \exp ( \frac{t}{b}  ) ^ c )  \Big\}$
    \newline
    Inverse Cumulative distribution Function:
\begin{equation} \nonumber
  t=   a \bigg[ \ln \bigg\{ 1 - ( \frac{1}{a b }) \ln (1-u) \bigg\} \bigg] ^ \frac{1}{c}
\end{equation}

\item[($iii$)]{Exponentiated Inverse Weibull Distribution   }
Notation used: $X \sim KW (a , b)  $ where $ a , c , t >0 , b \in \mathbb{R} $
    $SF$: $\bar{F}_{X}(t) = \{ 1 - \exp ( - a t^ {- c }   ) \} ^ b $
    Inverse Cumulative distribution Function:
\begin{equation} \nonumber
    t = \bigg[ \frac{(-a)}{\ln (1 - (1-u)^ \frac{1}{b})}  \bigg]^\frac{1}{c}
\end{equation}
\item[($iv$)]{Generalized Weibull Distribution (cf. Mudholkar, Srivastava, and Kollia (1996))}
    Notation used: $X \sim GW( a, b, c)  $ where $ a, c, b  >0 , 0 \leq t \leq (a c )^\frac{1}{b} $
    $SF$: $\bar{F}_{X}(t) = (1- a c t^ b )^\frac{1}{c}    $
   Inverse Cumulative distribution Function:
\begin{equation} \nonumber
    t = \bigg( \frac{1 - (1-u)^c }{a c }     \bigg) ^ \frac{1}{b}
\end{equation}
\item[($v$)]{Extended Weibull Distribution (cf. Ghitany, Al-Hussaini,
and Al-Jarallah (2005))}
    Notation used: $X \sim  EW( a,
b, c ) $ where $ a, b >0 , t \geq 0 $
    $SF$: $ \bar{F}_{X}(t) = \bigg[ \frac{a e^{- ( b t )^ c }}{1 - (1- a) (e^{ - (b t)^ c ) }} \bigg]   $
    Inverse Cumulative distribution Function:
\begin{equation} \nonumber
 t= (\frac{1}{b}) \bigg\{ \ln \bigg[ \frac{(2 a -1) + u(1-a )}{u} \bigg] \bigg\} ^ \frac{1}{c}
\end{equation}
\item[($vi$)]{ Generalized Power Weibull Distribution (cf. Nikulin and Haghighi (2006))}
    Notation used: $X \sim GPW( a, b, c)    $ where $ a, b, c >0 , t > 0 $

     $SF$: $\bar{F}_{X}(t) = \exp \{  1 - (1+ a t^b )^ \frac{1}{c}   \}$
    Inverse Cumulative distribution Function:
\begin{equation} \nonumber
    t= \bigg\{  \frac{[1- \ln (1-u) ]^c - 1 }{a} \bigg\} ^ \frac{1}{b}
\end{equation}
\item[($vii$)]{Odd Weibull Distribution (cf. Cooray (2006))}
    Notation used: $X \sim OW(a, b, c)  $ where $ a, b ,  c >0 , t \geq 0 $
     $SF$: $\bar{F}_{X}(t) = \{ 1 + (e^{ a t^b   } - 1) ^ c  \} ^ {-1}  $
    Inverse Cumulative distribution Function:
\begin{equation} \nonumber
   t = \bigg\{ \bigg( \frac{1}{a} \ln \bigg[  \bigg( \frac{u}{1-u} \bigg) ^ \frac{1}{c} + 1 \bigg] \bigg\} ^ \frac{1}{b}
\end{equation}

\end{enumerate}
\subsubsection{Four parameter Weibull Distribution}
\begin{enumerate}
    \item[($i$)] {Four Parameter Weibull (Kies, 1958) }
    Notation used: $X \sim KW(a, b, c, d)  $ where $ 0 \leq a < t < b , c, d > 0 $
    $SF$: $\bar{F}_{X}(t) = \exp { \bigg[   - c \bigg( \frac{t-a}{b-t}  \bigg) ^ d  \bigg] }$
    Inverse Cumulative distribution Function:
\begin{equation} \nonumber
 t= \frac{\bigg( \frac{b^d \ln (1-u) }{-c}  \bigg) ^ \frac{1}{d}
  + a }{    \bigg( \frac{ \ln (1-u) }{-c}  \bigg) ^ \frac{1}{d}
  + 1   }
\end{equation}
\end{enumerate}
\subsubsection{Five parameter Weibull Distribution}
\begin{enumerate}
    \item[$(i)$] {Exponentiated Kumaraswamy Weibull Distribution}
    Notation used: $X \sim  EKumW( a , b, c , d, e ) $ where $ a , b, c , d, e, t >0 ,  $
    \newline
    $SF$: $\bar{F}_{X}(t) =  1 - [ 1- \{ 1- ( 1 - e ^ { - d t ^ e}  ) ^ a  \} ^ b ] ^ c      $
    \newline
    Inverse Cumulative distribution Function:
\begin{equation} \nonumber
t=     \bigg\{- \big(\frac{\ln [ 1 - \{ 1 - [ 1 - ( 1-u ) ^ \frac{1}{c} ] ^ \frac{1}{b} \} ^ \frac{1}{a} ]}{d}\Big)  \bigg\} ^ \frac{1}{e}
\end{equation}
\end{enumerate}

\subsection{Weibull distributions having intractable inverse  CDFs}
There are several Weibull distributions whose inverse CDFs are not found in closed form. Here, we take up a few of those and get an approximation of inverse CDFs using Lambert $W$ function. For the sake of simple comprehension, we categorically list the different Weibull family on the basis of number of parameters involved in the distribution.
\subsubsection{Three parameter Weibull Distribution}
\begin{enumerate}
    \item[$(i)$] {\bf Three parameter Weibull Distribution (Lai et al., 2003)}\\
 A non-negative random variable $X$ having $SF$ of  three parameter Weibull distribution given by
$\bar{F}_{X}(t) = \exp (- a t^b e^{c t }),$  where $ a, b >0,
c, t \geq 0 $ is denoted by $X\sim WL(a,b,c).$ The failure rate of $X$ is given by $r(t) = a(b+ct)t^{b-1}e^{ct}$ which is increasing for $b \geq 1$ and is bathtub-shaped for $0 < b < 1$. \\
Since its inverse cumulative function is not in closed form, we approximate its  quantile function through Lambert $W$ function as given in following proposition.
\begin{p1}
If $X\sim WL(a,b,c)$ then its quantile function is approximated by \begin{equation}    Q(u) \approx \bigg( \frac{b}{c}   \bigg)   W \bigg[    \bigg( \frac{c}{b} \bigg) \bigg[  \bigg( \frac{-1}{a} \bigg) \ln (1-u) \bigg] ^ \frac{1}{b}      \bigg]. \nonumber
     \end{equation}
     \end{p1}
     {\bf Proof.}
     We substitute $ {F_X}(t) = u,$
i.e.,
\begin{equation}
     1 -  \exp (- a t^b e^{c t }    )    = u \nonumber
     \end{equation}
Or equivalently
  \begin{equation}    ( t^b e^{c t }    )    = \bigg( \frac{-1}{a} \bigg) \ln (1-u) \nonumber
  \end{equation}
  We arrange the left hand side in the form $f(x) \exp{f(x)}$, and get
   \begin{equation}       ( \frac{c t}{b} e^{   \frac{c t }{b}  }  )    = \bigg( \frac{c}{b} \bigg) \bigg[  \bigg( \frac{-1}{a} \bigg) \ln (1-u) \bigg] ^ \frac{1}{b}. \nonumber
   \end{equation}
 Now we make use of Lambert W function to obtain
     \begin{equation}      ( \frac{c t}{b}) = W \bigg[    \bigg( \frac{c}{b} \bigg) \bigg[  \bigg( \frac{-1}{a} \bigg) \ln (1-u) \bigg] ^ \frac{1}{b}      \bigg]. \nonumber        \end{equation}
Thus, the inverse $cdf$ is
     \begin{equation}    t = \bigg( \frac{b}{c}   \bigg)   W \bigg[    \bigg( \frac{c}{b} \bigg) \bigg[  \bigg( \frac{-1}{a} \bigg) \ln (1-u) \bigg] ^ \frac{1}{b}      \bigg].
     \end{equation}
     \item[($ii$)] {\bf Inverse Modified Weibull Distribution} \\
     If a non-negative random variable $X$ follows Inverse Modified Weibull Distribution, denoted by $X\sim IW(a,b,c)$ then its $SF$ is $$  \bar{F}_{X}(t) = 1 - \exp{(- {( \frac{a}{t}  )}^b \exp{( \frac{c}{t} )}  )},   $$ where $a,b,c\geq 0$ and $t\geq 0.$\\
     Since the inverse cumulative function of $IW$ distribution can not be found be in closed form, we approximate its  quantile function through Lambert $W$ function as stated in next proposition.
\begin{p1}
If $X\sim WL(a,b,c)$ then its quantile function is approximated by \begin{equation}
  Q(u) \approx \frac{c}{b}  \bigg(  W \bigg(   \frac{c}{b a}  ((-1) \log (u))^\frac{1}{b}   \bigg) \bigg)^ {-1} \nonumber
\end{equation}
\end{p1}
{\bf Proof.}
Taking $ {F_X}(t) = u,$
we  get
\begin{equation}
{\bigg( \frac{a}{t}  \bigg) }^b \exp{\bigg( \frac{c}{t} \bigg)}  =  (-1) \log (u).  \nonumber
\end{equation}
Further simplification gives
\begin{equation}
     \bigg( \frac{a}{t}  \bigg) \exp{\bigg( \frac{c}{b t} \bigg)}  =  ((-1) \log (u))^\frac{1}{b}.  \nonumber
\end{equation}
After rearranging,
\begin{equation}
\label{lw2}
 {\bigg( \frac{c}{b t}  \bigg)} \exp{\bigg( \frac{c}{b t} \bigg)}  =  \frac{c}{b a}  ((-1) \log (u))^\frac{1}{b} .
\end{equation}
For approximating by Lambert $W,$ we rewrite (\ref{lw2}) as
\begin{equation}
  \frac{c}{b t} = W \bigg(   \frac{c}{b a}  ((-1) \log (u))^\frac{1}{b}   \bigg).  \nonumber
\end{equation}
Thereby, the corresponding inverse $cdf$ is
\begin{equation}
  t =  \frac{c}{b}  \bigg(  W \bigg(   \frac{c}{b a}  ((-1) \log (u))^\frac{1}{b}   \bigg) \bigg)^ {-1}
\end{equation}

 \item[($iii$)] {\bf Three Parameter Weibull Distribution (cf. Xie and Lai, 1995)}
    We assume $X \sim XLW(a,b,c)   $ where $ t, c >0 , a \geq 0 , b> 1 $ if the corresponding  $SF$ is
$$\bar{F}_{X}(t) = \exp  ( -(at)^b - (at)^\frac{1}{b} - c t ).$$
To the best of our understanding we claim that the inverse $cdf$ of $X$ having $XLW$ distribution can not be approximated by Lambert $W$ function and it can be taken up as an open problem for future study.
\end{enumerate}
\subsubsection{Four parameter Weibull Distribution }
\begin{enumerate}
 \item[($i$)]{\bf Generalized Modified Weibull Distribution (cf. Carrasco,
Ortega, and Cordeiro (2008)):}
   The $SF$ of  Generalized Modified Weibull    Distribution  with $X\sim GMW( a, b, c, d) $ is  $$\bar{F}_{X}(t) =  1 - [ 1 - e ^ {- a t ^ c  e ^ { b t } } ]  ^ d      $$ where $ a, b, c, d, t  >0  $.
   \begin{p1}
   If $X\sim GMW (a,b,c,d)$ then an approximate of quantile function is given by
\begin{equation}   Q(u)\approx \bigg( \frac {c}{b} \bigg) W \bigg[   \bigg( \frac {b }{c} \bigg)
 \bigg[   \bigg( \frac{-1}{a } \bigg)    \ln (1 -   u ^ \frac{1}{d})  \bigg] ^ \frac{1}{c}  \bigg] \nonumber
 \end{equation}
   \end{p1}
 Here, $ {F_X}(t) = u,$
gives
 $ [ 1 - e ^ {-a t ^ c  e ^ { b t } } ]  ^ d   = u.$
 Equivalently,
     \begin{equation}
     t   e ^ {\frac {b t}{c} } =
 \bigg[   \bigg( \frac{-1}{a } \bigg)    \ln (1 -   u ^ \frac{1}{d})  \bigg] ^ \frac{1}{c}, \nonumber
 \end{equation}
 which can also be written as
      \begin{equation} \label{lw1}  \bigg( \frac {b t}{c} \bigg)    e ^ {\frac {b t}{c} } =  \bigg( \frac {b }{c} \bigg)
 \bigg[   \bigg( \frac{-1}{a } \bigg)    \ln (1 -   u ^ \frac{1}{d})  \bigg] ^ \frac{1}{c} .
 \end{equation}
 Now, applying the Lambert W function in (\ref{lw1}), we get
       \begin{equation}   \bigg( \frac {b t}{c} \bigg) = W \bigg[   \bigg( \frac {b }{c} \bigg)
 \bigg[   \bigg( \frac{-1}{a } \bigg)    \ln (1 -   u ^ \frac{1}{d})  \bigg] ^ \frac{1}{c}  \bigg]. \nonumber   \end{equation}
  Consequently,
        \begin{equation}   t = \bigg( \frac {c}{b} \bigg) W \bigg[   \bigg( \frac {b }{c} \bigg)
 \bigg[   \bigg( \frac{-1}{a } \bigg)    \ln (1 -   u ^ \frac{1}{d})  \bigg] ^ \frac{1}{c}  \bigg]
 \end{equation}
    \item[($ii$)]{\bf Shifted Modified Weibull Distribution:}
The $SF$ of a random variable $X$ following Shifted modified Weibull    Distribution is $$  \bar{F}_{X}(t) = \exp{(- {(a (t-d))}^b \exp{( c(t -d) )}  )},$$ where $a,b,c,d \geq 0.$ Here, we write $X\sim SMW(a,b,c,d).$
\begin{p1}
If $X\sim SMW(a,b,c,d)$ then its quantile function is approximated as \begin{equation}
Q(u) \approx d + \frac{b}{c}   W \bigg( \frac{c{((-1) \log(1-u)) }^\frac{1}{b} }{a b} \bigg)\nonumber
\end{equation}
\end{p1}
{\bf Proof.} To obtain its inverse cumulative function, we apply probability integral transform and substitute $ {F_X}(t) = u,$
and get
\begin{equation}
{(a (t-d))}^b \exp{( c(t -d) )} = (-1) \log(1-u),  \nonumber
\end{equation}
which is same as
\begin{equation}
 { a (t-d)} \exp{\big( \frac{c}{a}(t -d) \big)} = {((-1) \log(1-u) )}^\frac{1}{b}, \nonumber
\end{equation}
or equivalently
\begin{equation}
 {  (t-d)} \exp{\big( \frac{c}{a}(t -d) \big)} = \frac{{((-1) \log(1-u)) }^\frac{1}{b} }{a}. \nonumber
\end{equation}
After rearranging, it reduces to
\begin{equation}
 { \frac{c}{b} (t-d)} \exp{\big( \frac{c}{a}(t -d) \big)} = \frac{c{((-1) \log(1-u)) }^\frac{1}{b} }{a b}. \nonumber
\end{equation}
To express in terms of Lambert $W$, we get
\begin{equation}
 { \frac{c}{b} (t-d)}  = W \bigg[ \frac{c{((-1) \log(1-u)) }^\frac{1}{b} }{a b} \bigg],\nonumber
\end{equation}
i.e.,
\begin{equation}
 t = d + \frac{b}{c}   W \bigg( \frac{c{((-1) \log(1-u)) }^\frac{1}{b} }{a b} \bigg)
\end{equation}

\item[($iii$)]{\bf Additive Weibull Distribution} (cf. Xie and Lai (1996))
If $X$ follows additive Weibull Distribution then we denote $X \sim AW(a, b, c, d)  $ where $ a,b,c,d >0 , t \geq 0.$ The $SF$ of $X$ is $$\bar{F}_{X}(t) = \exp   ( -a t^b - c t^d ),$$ where $a,b,c,d \geq 0$ and $t\geq 0.$  To the best of our knowledge and effort, we believe that neither the inverse $cdf$ can be derived nor it can be approximated by Lambert $W$ function.
\item[($iv$)]{\bf Nadarajah and Kotz  Weibull Distribution} (cf. Nadarajah and Kotz (2005)) We denote the corresponding random variable as $X \sim NK(a,b,c,d)  $ where $ a, d >0 , b, c , t \geq 0 $ and its
    $SF$ is $$\bar{F}_{X}(t) = \exp  (-at^b(e^{ct^d}-1)).     $$
The inverse $cdf$ could not be retrieved using Lamber $W$ function for the said distribution.
\end{enumerate}
\subsubsection{Five parameter Weibull Distribution }
\begin{enumerate}
   \item[($i$)]{\bf Kumaraswamy Modified Weibull Distribution}\\
 The $SF$ of $X$ having Kumaraswamy modified Weibull  Distribution is $$   \bar{F}_{X}(t) =  [ 1 - (1- \exp{( - c t^d \exp{(\mu t)} )} )^a     ]^b,$$  where $ a,b,c, d, \mu >0 , t \geq 0. $ Accordingly, we write $X \sim  KMW( a,b,c,d,\mu ) $
\begin{p1}
If  $X \sim  KMW( a,b,c,d,\mu ) $ then its quantile function is approximated by
\begin{equation}
Q(u) \approx( \frac{d}{\mu})                  W (    \frac{\mu }{d} \{ (- \frac{1}{c})  \ln{ ( 1 - [ 1 - (1-u)^\frac{1}{b}  ]^\frac{1}{a}   )  }     \} ^ \frac{1}{d}    )  \nonumber
      \end{equation}
\end{p1}
{\bf Proof.} To obtain its inverse cumulative function, we apply probability integral transform and substitute $ {F_X}(t) = u,$
i.e.,
\begin{equation}
 1 - [ 1 - (1- \exp{( - c t^d \exp{(\mu t)} )} )^a     ]^b = u   \nonumber\end{equation}
It follows that
\begin{equation}
 - c t^d \exp{(\mu t)} ) = \ln{ ( 1 - [ 1 - (1-u)^\frac{1}{b}  ]^\frac{1}{a}   )  }   \nonumber
 \end{equation}
After some algebraic manipulations, we obtain the required $f(x) \exp{f(x)}$ form to put the Lambert W function into use
\begin{equation}
 \frac{\mu t}{d} \exp{( \frac{\mu t}{d}  )}  = \frac{\mu }{d} \{ -( \frac{1}{c})  \ln{ ( 1 - [ 1 - (1-u)^\frac{1}{b}  ]^\frac{1}{a}   )  }     \} ^ \frac{1}{d} \nonumber
 \end{equation}
Now, we are ready to use the Lambert W function
\begin{equation}
    \frac{\mu t}{d} =                     W (    \frac{\mu }{d} \{ (- \frac{1}{c})  \ln{ ( 1 - [ 1 - (1-u)^\frac{1}{b}  ]^\frac{1}{a}   )  }     \} ^ \frac{1}{d}    ) \nonumber \end{equation}

Consequently, we get the inverse $cdf$ as
\begin{equation}
      t =   ( \frac{d}{\mu})                  W (    \frac{\mu }{d} \{ (- \frac{1}{c})  \ln{ ( 1 - [ 1 - (1-u)^\frac{1}{b}  ]^\frac{1}{a}   )  }     \} ^ \frac{1}{d}    )
      \end{equation}
             \item[($ii$)]{\bf Five parameter Weibull Distribution, (Phani, 1987)}
    Notation used: $X \sim W_5 (a,b, c, d, e )  $ where $ 0 \leq a < t < b , c , d ,  e >   0 $
    $SF$: $\bar{F}_{X}(t) = \exp \bigg[    -c \frac{(t-a)^ d}{(b-t)^e}        \bigg]   $
\end{enumerate}

\section{Simulation of some non-Weibull distributions }
In this section, we take up some statistical distributions not belonging to Weibull family whose inverse cumulative distribution  are not in closed form. As a result, their $icdf$ simulation study in the line of simulation plays a crucial role in the sampling procedure.\\
In particular, we focus on lifetime distributions, viz., modified Log Logistic, Gompertz Makeham, modified Power Lomax, modified Pareto IV, modified Lognormal distributions and see
\begin{enumerate}
    \item[($i$)]{{\bf Modified Log Logistic Distribution}}\\
A random variable $X$ which has Modified Log Logistic     Distribution has $SF,$ $$\bar{F}_{X}(t) = [1 + ( a t )^ b e ^ { c t }] ^ {-1},$$ where $a,b,c\geq 0,$ and $t\geq 0.$ We write $X\sim MLL(a,b,c).$
\begin{p1}
If $X\sim MLL(a,b,c)$ then its approximated quantile function is \begin{equation} Q(u) \approx  ( \frac{b}{c}   )       W [ \frac{c  }{ a  b}
[  (1- u) ^ {-1} - 1 ]^ \frac{1}{b} ] \nonumber
\end{equation}
\end{p1}
{\bf Proof.} To obtain its inverse cumulative function, we apply probability integral transform and substitute $ {F_X}(t) = u,$
i.e.,
 \begin{equation}
   1 - [1 + ( a t )^ b e ^ { c t }  ] ^ {-1}  = u  \nonumber
   \end{equation}
After some algebraic manipulations, we obtain
\begin{equation} \label{qu}  [ ( \frac{c t }{b} ) e ^ { \frac{c t }{b}  }]   = \frac{c  }{ a  b}
[  (1- u) ^ {-1} - 1 ]^ \frac{1}{b}
\end{equation}
We first  rearrange left hand side of (\ref{qu}) in the form $f(x) \exp{f(x)} $, to apply Lambert W function and obtain
\begin{equation}    [ ( \frac{c t }{b} ) ]   = W [ \frac{c  }{ a  b}
[  (1- u) ^ {-1} - 1 ]^ \frac{1}{b}. \nonumber
\end{equation}
So we have,
\begin{equation}   t =  ( \frac{b}{c}   )       W [ \frac{c  }{ a  b}
[  (1- u) ^ {-1} - 1 ]^ \frac{1}{b} ]
\end{equation}
     \item[($ii$)]{\bf Gompertz Makeham Distribution}:
The $SF$ of Gompertz Makeham is
$$\bar{F}_{X}(t) =   \exp{(-  a t - \frac{b}{c}  [\exp{( c t   )} - 1  ] )}, $$ where $a,b,c\geq 0,$ and $t\geq 0.$ We write $X\sim GM(a,b,c)$
\begin{p1}
If $X\sim GM(a,b,c)$ then its quantile function is approximated by $$ Q(u) \approx  \ln {( \{  (\frac{a}{b})     W[  (\frac{b}{a})      \exp{(\frac{b  - c \ln{(1-u)}}{a })}  ]    \} ^ \frac{1}{c}  )}.$$

\end{p1}
{\bf Proof.} We now proceed to obtain its inverse cumulative function, and substitute $ {F_X}(t) = u,$
i.e.,
$$  1- \exp{(-  a t - \frac{b}{c}  \{\exp{( c t   )} - 1  \} )}  =u,$$ or equivalently
\begin{equation}
\label{gomp}
 a t + \frac{b}{c}  [\exp{( c t   )} - 1  ]  =  - \ln{(1-u)}.
\end{equation}  We noted that
(\ref{gomp}) does not lead us to an explicit relation of $t$ in terms of $u$, thereby, in the next few lines, we intend to get a closed form by use of Lambert $W$. To do so, we rewrite (\ref{gomp}) as
\begin{equation}
\label{gomp1} a \ln{e^t} + \frac{b}{c} (e^t)^c = \frac{b}{c} - \ln{(1-u)}.
\end{equation}
Taking exponential on both sides of (\ref{gomp1}), we get
\begin{equation}
\label{gomp2}
\exp{( a \ln{e^t} + \frac{b}{c} (e^t)^c)}   = \exp{(\frac{b}{c} - \ln{(1-u)})}.
\end{equation}
(\ref{gomp2}) is identical to
\begin{equation}
\label{gomp3}
\exp{( \ln({\exp(t)})^a )} \exp{( \frac{b}{c} (\exp(t))^c )}  = \exp{(\frac{b}{c} - \ln{(1-u)})}.
\end{equation}
Rearranging (\ref{gomp3}), we have
$$ ({\exp(t)})^a  \exp{\bigg( \frac{b}{c} (\exp(t))^c \bigg)}  = \exp{\bigg(\frac{b}{c} - \ln{(1-u)}\bigg)}, $$
or correspondingly,
\begin{equation} \label{gomp4} ({\exp(t)})  \exp{( \frac{b}{a c} (\exp(t))^c )}  = \exp{\bigg(\frac{b  - c \ln{(1-u)}}{a c}\bigg)}.\end{equation}
We rewrite (\ref{gomp4}) as
\begin{equation}  \label{gomp5} ({\exp(t)})^c  \exp{\big( \frac{b}{a } (e^t)^c )}  = \exp{(\frac{b  - c \ln{(1-u)}}{a }\big)}.      \end{equation}
We now arrange (\ref{gomp5}) in the form $f(\cdot)\exp{\big(f(\cdot)\big)},$ to bring in Lambert $W$ and subsequently obtain
$$  \big(\frac{b}{a}\big)     \big({\exp(t)}\big)^c  \exp{( \frac{b}{a } (\exp(t))^c )}  =   (\frac{b}{a})      \exp{(\frac{b  - c \ln{(1-u)}}{a })}           $$
Thus,
$$  (\frac{b}{a})     ({\exp(t)})^c   = W[  (\frac{b}{a})      \exp{(\frac{b  - c \ln{(1-u)}}{a })}  ].$$
After further simplification, we aim to solve for $t,$ giving
$${\exp(t)}   =   \{  (\frac{a}{b})     W[  (\frac{b}{a})      \exp{(\frac{b  - c \ln{(1-u)}}{a })}  ]    \} ^ \frac{1}{c} $$
and eventually get
\begin{equation}t  =  \ln {( \{  (\frac{a}{b})     W[  (\frac{b}{a})      \exp{(\frac{b  - c \ln{(1-u)}}{a })}  ]    \} ^ \frac{1}{c}  )}.\nonumber\end{equation}
\item[($iii$)]{\bf Modified Power Lomax Distribution:} The $SF$ of a random variable $X$ having Modified Power Lomax     Distribution is $$\bar{F}_{X}(t) =  [ 1 + ( a t ) ^ b e ^ {( c t )} ] ^ { - d}, $$ where $a,b,c,d\geq 0$ and $t\geq 0.$ We write $X\sim MPL(a,b,c,d).$
\begin{p1}
If $X \sim MPL(a,b,c,d)$ then its quantile function is approximated as \begin{equation}    Q(u) \approx ( \frac{b }{c} )   W [ ( \frac{c }{a  b} )   [( 1 - u ) ^ { \frac{-1}{ d}} - 1 ] ^ \frac{1}{b} ] \nonumber
\end{equation}
\end{p1}
{\bf Proof.}
 With an objective to obtain its inverse cumulative function, we substitute $ {F_X}(t) = u,$ and get
 \begin{equation}   1 -  [ 1 + ( a t ) ^ b e ^ {( c t )} ] ^ { - d} = u.  \nonumber
\end{equation}
 It follows that
\begin{equation}    [ 1 + ( a t ) ^ b e ^ {( c t )} ] ^ { - d} = (1 - u).  \nonumber
\end{equation}
Then we have
\begin{equation}     ( \frac{c t}{b} ) e ^ {  \frac{( c t )}{b}}  = ( \frac{c }{a  b} )   [( 1 - u ) ^ { \frac{-1}{ d}} - 1 ] ^ \frac{1}{b}\nonumber
\end{equation}
Now using the Lambert W function
\begin{equation}   [  ( \frac{c t}{b} ) ]  = W [ ( \frac{c }{a  b} )   [( 1 - u ) ^ { \frac{-1}{ d}} - 1 ] ^ \frac{1}{b} ] \nonumber
\end{equation}
Consequently,
\begin{equation}    t  = ( \frac{b }{c} )   W [ ( \frac{c }{a  b} )   [( 1 - u ) ^ { \frac{-1}{ d}} - 1 ] ^ \frac{1}{b} ]
\end{equation}
      \item[($iv$)]{\bf Modified Pareto IV Distribution:}
The $SF$ of    Modified Pareto IV  Distribution is $$\bar{F}_{X}(t) =  [ 1 + ( a (t-\mu) ) ^ {\frac{1}{b}} e ^ {( c (t-\mu) )} ] ^ { - d}, $$ where $a,b,c,d,\mu\geq 0,$ and $t\geq 0.$ We write $X\sim MPIV(a,b,c,d,\mu)$
    \begin{p1}
If $X\sim MPIV(a,b,c,d,\mu)$ then its quantile function is approximated by \begin{equation}Q(u) \approx ( \frac{1}{c b})  W [
[ ( 1 - u ) ^ { \frac{-1}{ d}} - 1 ] ^ b ] + \mu.\nonumber
\end{equation}
 \end{p1}
{\bf Proof.}
Here, $ {F_X}(t) = u,$
gives
 \begin{equation}  1 -  [ 1 + ( a (t-\mu) ) ^{\frac{1}{b}} e ^ {( c (t-\mu) )} ] ^ { - \beta} = u.  \nonumber
 \end{equation}
Equivalently, we write
\begin{equation}     ( a (t-\mu) ) e ^ {( c (t-\mu) )}   =
[ ( 1 - u ) ^ { \frac{-1}{ d}} - 1 ] ^ b. \nonumber
\end{equation}
It follows from the use of the Lambert W function that
\begin{equation}   [  ( a (t-\mu) ) ]  = W [
[ ( 1 - u ) ^ { \frac{-1}{ d}} - 1 ] ^ b ]\nonumber
\end{equation}
The required approximated solution is obtained as
\begin{equation}     t    = ( \frac{1}{c b})  W [
[ ( 1 - u ) ^ { \frac{-1}{ d}} - 1 ] ^ b ] + \mu.
\end{equation}
\item[($v$)]{\bf Modified Lognormal Distribution:}
The $SF$ of Modified Lognormal is $$\bar{F}_{X}(t) = 1 - \Phi ( \frac{ \ln{ (a t)^b} \exp{ (c t) } - d    }{  \mu  } ), $$ where $ \Phi(t) = \frac{1}{\sqrt{2 \pi}     }  \int_{0} ^ {t} \exp {(\frac{-t^2}{2})}dt.$ We write $X\sim MLN(a,b,c,d,\mu).$\\
The next proposition gives an approximated quantile function of modified lognormal distribution.
\begin{p1}
If $X\sim MLN(a,b,c,d,\mu)$ then its quantile function is approximated by \begin{equation}
Q(u) \approx ( \frac{b}{c}  )  W \bigg( ( \frac{c}{a b})  ( \exp ( \mu \Phi^{-1} (u) + d) )^\frac{1}{b} \bigg).\nonumber
  \end{equation}
\end{p1}
{\bf Proof.}
We substitute $ {F_X}(t) = u,$ gives
 \begin{equation}   \frac{ \ln{ (a t)^b } \exp{ (c x) } - d    }{  \mu  }  =  \Phi^{-1} (u)\nonumber
 \end{equation}
 Further analysis gives,
  \begin{equation}
 ( a t)^b  \exp (c t)   = \exp ( \mu \Phi^{-1} (u) + d ),\nonumber
  \end{equation}
 which reduces to \begin{equation}
 ( a t) \exp (\frac{c t}{b})   =   ( \exp ( \mu \Phi^{-1} (u) + d ) )^\frac{1}{b}.\nonumber
  \end{equation}
  Now to bring in Lambert $W$ function, we make necessary simplification and get
\begin{equation}
 ( \frac{c t}{b}) \exp (\frac{c t}{b})   = ( \frac{c }{a b })  ( \exp ( \mu \Phi^{-1} (u) + d ) )^\frac{1}{b}.\nonumber
  \end{equation}
  Thus,
\begin{equation}
 ( \frac{c t}{b}) = W \bigg( ( \frac{c }{a b })  ( \exp ( \mu \Phi^{-1} (u) + d ) )^\frac{1}{b} \bigg),\nonumber
  \end{equation}
and the inverse $cdf$ is approximated as
\begin{equation}
t =  ( \frac{b}{c}  )  W \bigg( ( \frac{c}{a b})  ( \exp (\mu \Phi^{-1} (u) + d) )^\frac{1}{b} \bigg).
  \end{equation}
\end{enumerate}

    \end{document}